\documentclass[aps,prl,twocolumn,superscriptaddress]{revtex4-1}

\usepackage[pdftex]{graphicx}

\begin{document}

\title{How robust the global measurements of London penetration depth are: the case of Fe(Te$_{1-x}$,Se$_{x}$)}

\author{K.~Cho}
\affiliation{The Ames Laboratory, Ames, IA 50011, USA}

\author{H.~Kim}
\affiliation{The Ames Laboratory, Ames, IA 50011, USA}
\affiliation{Department of Physics \& Astronomy, Iowa State University, Ames, IA 50011, USA}

\author{M.~A.~Tanatar}
\affiliation{The Ames Laboratory, Ames, IA 50011, USA}

\author{J.~Hu}
\affiliation{Department of Physics and Engineering Physics, Tulane University, New Orleans, LA 70118}

\author{B.~Qian}
\affiliation{Department of Physics and Engineering Physics, Tulane University, New Orleans, LA 70118}

\author{Z.~Q.~Mao}
\affiliation{Department of Physics and Engineering Physics, Tulane University, New Orleans, LA 70118}

\author{R.~Prozorov}
\email[Corresponding author: ]{prozorov@ameslab.gov}
\affiliation{The Ames Laboratory, Ames, IA 50011, USA}
\affiliation{Department of Physics \& Astronomy, Iowa State University, Ames, IA 50011, USA}

\date{4 July 2011}

\begin{abstract}
We report tunnel diode resonator measurements of in-plane London penetration depth, $\lambda(T)$, in optimally-doped single crystals of Fe(Te$_{0.58}$Se$_{0.42}$) with $T_c\sim$ 14.8 K. Systematic measurements were carried out for six samples with different size and surface roughness. The power-law behavior, $\Delta \lambda(T) = A T^n$ was found for all samples with the average exponent $n_{avg} = 2.3 \pm 0.1$ and the pre-factor $A_{avg} = 1.0 \pm 0.2$ nm/K$^{2.3}$. The average superfluid density is well described by the self-consistent two-gap $\gamma$ model resulting in $\Delta_{_{I}}(0)$/$k_B T_c$ = 1.93 and $\Delta_{_{II}}(0)$/$k_B T_c$ = 0.9. By analyzing the data obtained on samples of different size and deliberately introduced surface roughness, it is concluded that the calibration procedure used to obtain $\lambda(T)$ from the measured TDR frequency shift is quite robust and the uncertainty in sample dimensions and the nature of surface roughness play only minor role. The exponent $n$, directly related to the superconducting gap structure, remains virtually unchanged. The calibration - dependent pre-factor $A$ shows some variation, but stays within reasonable margin ruling out some recent suggestions that surface conditions can significantly affect the results. Our results confirm that precision global measurements provide the most objective information regarding temperature - dependent London penetration depth.
\end{abstract}



\maketitle

The crystals of iron-based superconductors are micaceous and soft, which prompts questions related to sample structure and homogeneity. Many measurements assume perfect uniformity of the bulk properties and perfectly flat surface. This is especially important for the measurements of London penetration depth, $\lambda(T)$, which is one of the key quantities that can be measured with great precision providing valuable information on the quasiparticle excitations, thus the superconducting gap structure. When the entire sample is measured, several factors may affect the outcome. Size and shape enter the calibration factors that is used to convert raw data into the actual length. Chemical inhomogeneity may affect the measurements as an intrinsic source. One recent example of the experimental controversy is the comparison of $\lambda(T)$ estimated from the tunnel-diode resonator (TDR) and first critical field measurements in single crystals of Fe(Se,Te) \cite{Klein}. In order to reconcile two data sets the authors had to assume that the effective sample dimension, $R$, in the TDR calibration should be five times larger than the calculated from the sample dimensions, presumably due to surface roughness.

\begin{figure}[tb]
\includegraphics[width=8.5cm]{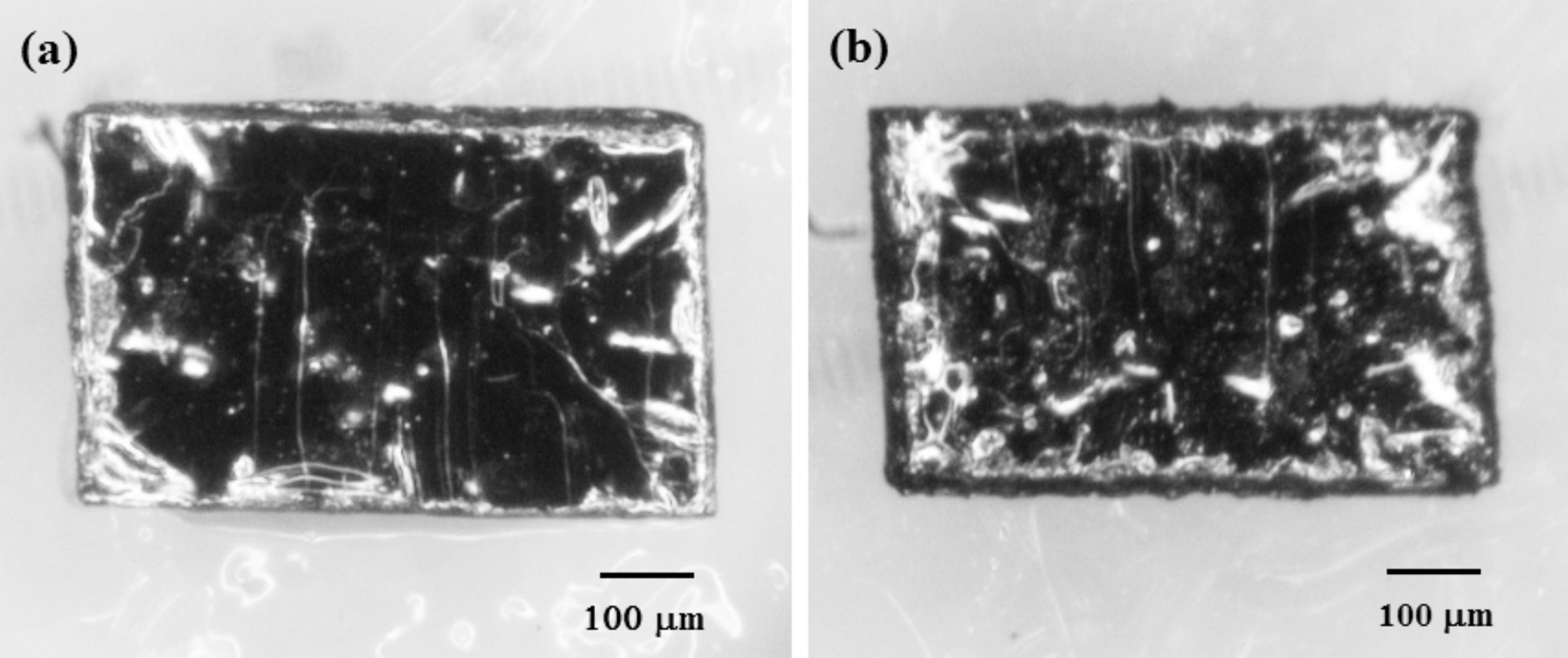}
\caption{\label{fig.1} (Color online) (a) Sample 2 - before and after (b) (Sample 2-R) deliberate roughening, see Table~\ref{tab.1}}
\end{figure}

In order to test the effects of sample size, shape and surface roughness we measured London penetration depth in optimally doped Fe(Te$_{0.58}$Se$_{0.42}$) with $T_c \sim 14.8$ K using a TDR technique. A series of measurements were carried out on parts of the same sample with different sizes and surface roughness. We find a robust power-law variation, $\lambda(T) \sim A T^n$ with $n$ = 2.3 $\pm$ 0.1 among all measurements indicating intrinsic behavior. For the purpose of investigating how surface roughness affects $\Delta \lambda (T)$, the edges of one of the samples were deliberately damaged by a blade, see Fig.~\ref{fig.1}. Measurements of this sample before and after damaging the edges show no significant change of the pre-factor $A$ or the exponent $n$. These results suggest that London penetration depth measured by a TDR technique represents an intrinsic quantity provided that the samples exhibit homogeneous superconductivity. We emphasize that it is the exponent $n$ that is most important quantity to characterize the gap structure and possible pair-breaking effects.

The iron chalcogenide family of superconductors Fe$Ch$ ($Ch$ = Se/Te/S) was discovered in 2008 \cite{Hsu} shortly after the discovery of the first iron based superconductor La(O$_{1-x}$F$_x$)FeAs \cite{Kamihara}. The structure of this family is simpler than those of other Fe-based superconductors (FeSC) \cite{Rotter, Ogino, Zhu, Wang}. The square planar sheets of Fe ions are alternated by the distorted tetrahedra of Chalcogen ions. Even with simpler structure, the electronic structure of Fe$Ch$ is similar to other FeSCs. Superconductivity in Fe$Ch$ appears upon substitution of Te with Se or S \cite{Yeh, Mizuguchi}, or control of the amount of excess Fe \cite{Hsu, Bendele}. The superconducting transition temperature T$_c$ is lower than those in most other FeSCs. In Fe$_y$(Te$_{1-x}$Se$_{x}$) at optimal doping, $T_c$ reaches a value of $\sim$ 14~K and $\sim$ 36~K under high pressure \cite{Hsu, Fang, Liu2009,Khasanov, Margadonna, Sales}. A series of theoretical and experimental studies suggest that superconductivity in Fe$Ch$ could be magnetically mediated following the spin-fluctuation picture \cite{Subedi, Xia, Bao, Qiu}.

The pairing mechanism of Fe$Ch$ has been explored from various measurements \cite{Kotegawa, Dong}. In nuclear magnetic resonance (NMR) study of polycrystalline FeSe ($T_c$ = 8 K), the unconventional pairing was suggested from the absence of the NMR coherence peak in conjunction with the power-law temperature dependence of the spin-relaxation rate, $1/T_{1} \sim T^{3}$ \cite{Kotegawa}. This can be explained with the unconventional order parameter such as a fully gapped s$_{\pm}$ or a nodal gap pairing.
In thermal conductivity measurements on FeSe$_x$ single crystals, the multi-gap nodeless superconductivity was suggested \cite{Dong}.

The London penetration depth in Fe$Ch$ has been measured by using different methods \cite{Khasanov2, Kim2, Serafin,Klein}. The result from muon spin rotation ($\mu$SR) in FeSe$_x$ is consistent with either anisotropic s-wave or a two-gap extended s-wave pairing \cite{Khasanov2}. Radio-frequency TDR measurements of $\lambda_{ab}(T)$ in Fe$_{y}$(Te$_{1-x}$Se$_x$) ($x$ = 0.37 \cite{Kim2}, $x$ = 0.44 \cite{Serafin}, and $x$ =0.45 \cite{Klein}) by three different groups found similar power-law exponent $n$ and some variation in the pre-factor $A$: $n \sim 2.0$, $A \sim $3.7 nm/K$^{n}$  \cite{Kim2}, $n \sim 2.2$, $A \sim $ 0.9 nm/K$^n$ \cite{Serafin} and $n \sim 2.0$, $A \sim$ 4 nm/K$^{n}$ \cite{Klein}. The full-range temperature behavior was also very similar. However, as mentioned above, an attempt to compare the TDR results with $\lambda(T)$ estimated from the measured first critical field, $H_{c1}$ using Ginzburg-Landau formula (valid only at $T_c$) resulted in a large discrepancy forcing the Authors to suggest a very different effective dimension of $R$ = 70 $\mu$m as compared to the one calculated from the sample dimensions, $R$ = 14 $\mu$m \cite{Klein}.

\begin{table}
\caption{List of samples and their physical constraints. $R$ is the effective dimension calculated from the dimensions of a sample \cite{Prozorov}.}
\label{tab.1}
\begin{center}
\begin{tabular}{|l|c|c|c|}
\hline
Sample & Dimensions ($\mu$m$^3$) & \textit{R} ($\mu$m) & Edge condition\\
\hline
1     & 712 $\times$ 488 $\times$ 40 & 33.5 & clean cut\\
1-A   & 491 $\times$ 270 $\times$ 40 & 22.6 & clean cut\\
1-B   & 489 $\times$ 443 $\times$ 40 & 28.4 & clean cut\\
2     & 702 $\times$ 455 $\times$ 40 & 32.8 & clean cut\\
2-R   & 680 $\times$ 430 $\times$ 40 & 31.5 & rough edge\\
2-R-C & 527 $\times$ 421 $\times$ 40 & 28.6 & clean cut\\
\hline
\end{tabular}
\end{center}
\end{table}

Single crystals of Fe(Te$_{1-x}$Se$_{x}$) were synthesized using a flux method as reported before \cite{Liu2009}. The samples studied in this work had composition of Fe : Te : Se = 1.00 $\pm$ 0.01 : 0.58 $\pm$ 0.01 : 0.42 $\pm$ 0.01 determined from the energy dispersive x-ray spectroscopy (EDXS). Initially two samples 1 and 2 were prepared by careful cleaving and cutting processes. Then Sample 1 was split into Samples 1-A and 1-B. To investigate the influence of the edge roughness, deliberate damage was done to Sample 2 by a razor blade. The original (Sample 2) and roughened ones (Sample 2-R) are shown in Fig.~\ref{fig.1}. To produce heaviest roughness possible, the damaging process was repeated multiple times resulting in a loss of about 10 \% of volume. After the measurements, the rough surface of Sample 2-R was cut again as clean as possible to remove the roughness and the cleaned one was labeled as 2-R-C. The penetration depths was measured at each stage of the described procedures. The dimensions and edge condition of samples are summarized in Table~\ref{tab.1}.

\begin{figure}[tb]
\includegraphics[width=8.5cm]{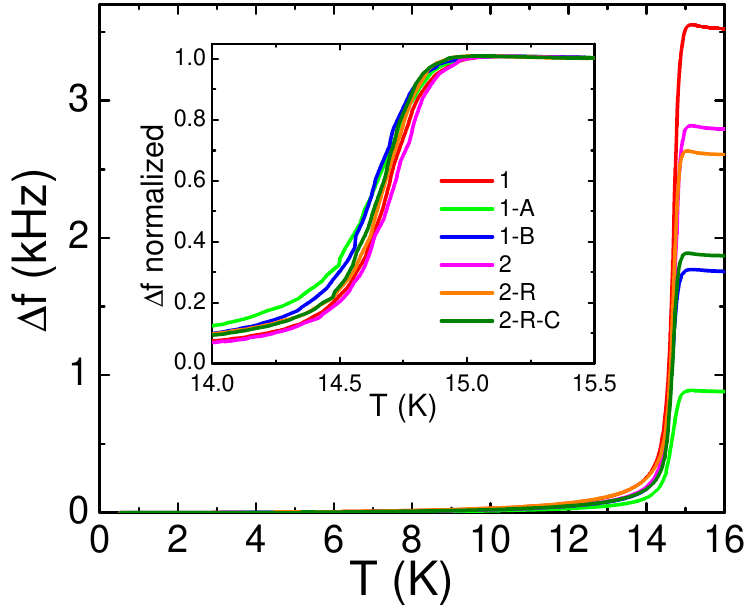}
\caption{\label{fig.2} (Color online) Temperature dependent frequency shift, $\Delta f(T)$ for each sample listed in Table~\ref{tab.1}. $T_c$ is consistently determined to be 14.8 K by using onset curves at the phase transition. Inset : $\Delta f(T)$ normalized based on the value at 16 K. The high temperature region is zoomed in showing the phase transition.}
\end{figure}

The in-plane London penetration depth, $\lambda (T)$, was measured using a self-oscillating tunnel-diode resonator (TDR) technique \cite{Prozorov, Van Degrift, Prozorov2}. A sample under study is mounted on a sapphire rod and inserted into an inductor coil of a LC tank circuit. To measure the in-plane penetration depth, the sample is placed with its c-axis along the direction of ac-field ($H_{ac}$) induced by the inductor coil. Since $H_{ac}$ $\sim$ 20 mOe is weak enough ($\ll H_{c1}$), the sample is in the Meissner state, so its magnetic response is determined by the London penetration depth. The frequency shift in TDR, $\Delta f \equiv f(T)-f_0$, is used to obtain the magnetic susceptibility $\chi(T)$ from $\Delta f = -G 4 \pi \chi (T)$. Here $f_0 = 1/2 \pi \sqrt{LC} \sim$ 14 MHz is the resonant frequency of an empty resonator, $G = f_0 V_s / 2 V_c (1-N)$ is a geometric factor defined by the coil and sample volumes, $V_c$ and $V_s$, and $N$ is the demagnetization factor. Calibration constant $G$ is directly measured by pulling the sample out of the coil at the lowest temperature. In the Meissner state, $\lambda$ can be obtained from the magnetic susceptibility, $\chi$, following the relation $4 \pi \chi = (\lambda / R)$tanh$(R/\lambda)-1$ where $R$ is the effective dimension of the sample \cite{Prozorov}.

\begin{figure}[tb]
\includegraphics[width=8.5cm]{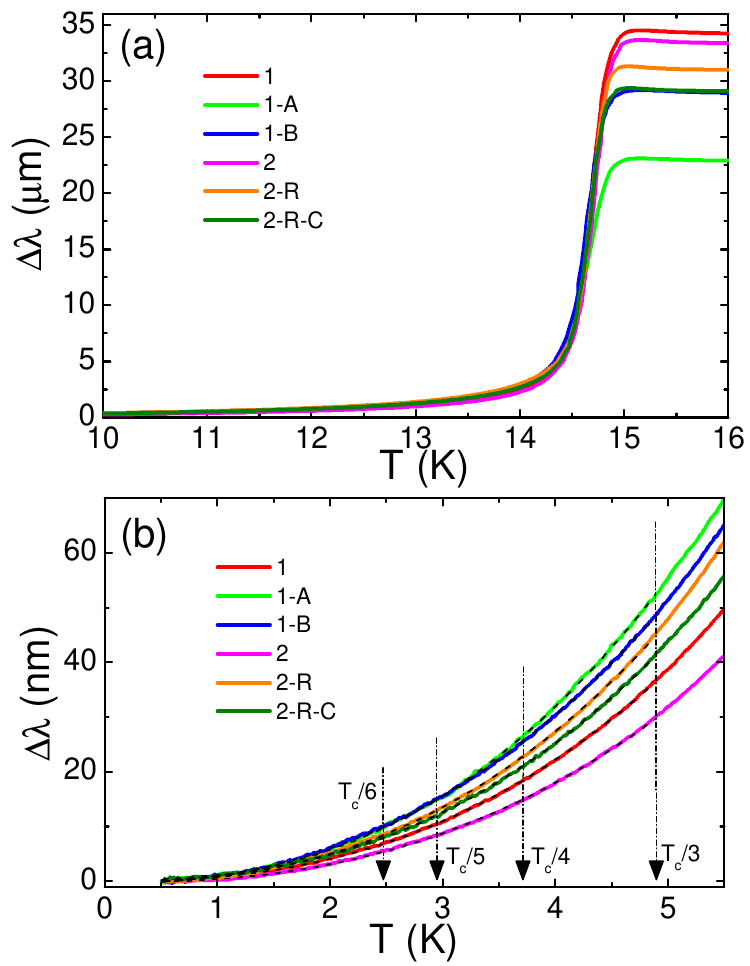}
\caption{\label{fig.3} (Color online) Temperature dependent penetration depth, $\Delta \lambda (T)$ for each sample. (a) High temperature region and (b) Low temperature region of $\Delta \lambda (T)$. The arrows indicate the upper-limits of fitting. The dashed lines are representative fits for each sample conducted up to $T_c$/3. All the fitting results are summarized in Fig.~\ref{fig.4}}
\end{figure}

Figure~\ref{fig.2} shows the raw data, temperature dependent frequency shift, $\Delta f (T)$, for samples listed in Table~\ref{tab.1}. It is obvious that all the samples have quite sharp and well overlaping transition curves with the width of the superconducting transition of about $\sim$ 0.3 K, so it seems that intrinsic superconducting properties were consistently reproducible between the samples (all samples come from the same large piece).

\begin{figure}[tb]
\includegraphics[width=8.5cm]{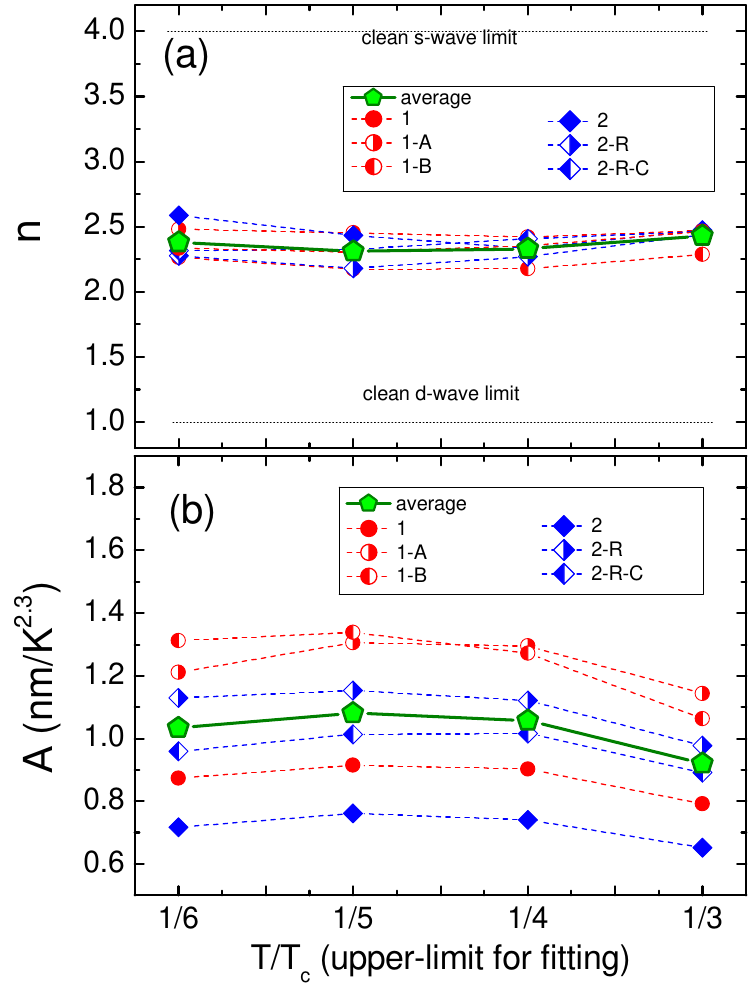}
\caption{\label{fig.4} (Color online) Results of the power-law fits to $\Delta \lambda (T)$ = $A T^n$. Four different upper-limits, indoicated by arrows in Fig.~\ref{fig.3}(b), were used. Panel (a): the exponent $n$, obtained by keeping $A$ and $n$ as free parameters. Panel (b): the pre-factor $A$, obtained at a fixed $n_{avg}=2.3$, which is the average among all seven samples shown in Panel (a).}
\end{figure}

Above $T_c$ the signal, $\Delta f (T)$, saturates due to two possible reasons. One, so-called sample-size-limited regime, is due to the size of a sample, which affects the $\tanh{R/\lambda}$ factor in the expression for the differential susceptibility. Another reason for the saturation, so-called skin-depth limited, occurs if the penetration depth becomes limited by the normal state skin depth. At $T_c$, the estimate value of the skin depth is about twice of that estimated from our calibration for $\lambda(T)$ \cite{Hardy93}. Thus, the samples under current study is in sample-size-limited regime, so their penetration depth shifts, $\Delta \lambda (T)$, are calculated from $\Delta f(T)$ in Fig.~\ref{fig.3} following the calibration procedure explained above. Figure~\ref{fig.3}(a) shows full temperature range which reveals sharp phase transitions at 14.8 K. The low temperature behavior of $\Delta \lambda$  is shown in Fig.~\ref{fig.3}(b). From these calibrated values we see that the penetration length in the normal state is comparable to the values of $R$, so we are in a sample-size limited regime. Therefore, the total frequency shift between the lowest temperature ($\sim$ 500 mK) and the normal state temperature ($T > T_c$), $\Delta f_{total}$, is related to the sample dimension. For instance, the largest Sample 1 has the biggest $\Delta f_{total}$ and the smallest Sample 1-A the smallest $\Delta f_{total}$. The inset shows $\Delta f(T)$ normalized by the value at 16 K.

For quantitative analysis of the low temperature behavior, a power-law fit, $\Delta \lambda (T)$ = $A T^n$ was performed for all six samples. To check how robust is the power law, in each case, four different upper temperature limits for the fit were used $T_c$/3, $T_c$/4, $T_c$/5 and $T_c$/6, shown schematically by arrows in Fig.~\ref{fig.3}(b). The results of the fitting are summarized in Fig.~\ref{fig.4}. Clearly, the fit coefficients remain fairly constant with small (and expected) deviations at the lowest and the highest limits. Below we discuss the results for $T_c/4$ chosen as the upper limit for the fit.

As shown in Fig.~\ref{fig.4} (a), in general, the exponent $n$ is rather steady among all cases even though there are small variations. From the best fits with $T/T_c$ upper limit, the average exponent $n_{avg}=2.3\pm 0.1$ is quite comparable to the previous reports of $n=$2.1 \cite{Kim2}, 2.2 \cite{Serafin} and 2.0 \cite{Klein}. In conjunction with the superfluid density analysis discussed later, the exponent $n=$ 2.3 can be explained by the nodeless two-gap pairing symmetry with strong pair breaking effect. The behavior of the pre-factor $A$ is summarized in Fig.~\ref{fig.4}(b). Clearly, there is a very weak dependence on the fitting range, but there is a more substantial change between the samples. The average over all samples pre-factor $A_{avg} = 1.0 \pm 0.2$ nm/K$^{2.3}$ is similar to other reports \cite{Kim2,Serafin}.

Let us now try to understand the observed variation of $A$ for different samples. We choose the values obtained from the fitting range up to $T_c/4$ where the fit quality was the best for all samples. We started by measuring Sample 1 that showed $A=$ 0.9 nm/K$^{2.3}$. After then Sample 1 was cut into Samples 1-A and 1-B. Every attempt was made to cut as clean as possible, but the pre-factor $A$ has increased \emph{for both samples} 1-A and 1-B to 1.3 nm/K$^{2.3}$. Although this change is not too significant, this result suggests that the effect of the rough edges becomes more pronounced as the sample size decreases. To estimate the effect of roughness itself when the size of a sample remains the same, Sample 2-R was made out of Sample 2 by deliberately damaging the edge with a razor blade. However, it turns out that the increment of $A$ was only by 0.4 nm/K$^{2.3}$ from 0.7 nm/K$^{2.3}$ for Sample 2 to 1.1 nm/K$^{2.3}$ for Sample 2-R. This increment is similar to that between Sample 1 and two Samples 1-A and 1-B. This means that the intentional roughness created by coarse roughening a sample doesn't significantly affect the result and it also means that the effective sample dimension $R$ used to calibrate the TDR data (so that it directly affects the pre-factor $A$) is very close to the calculated value. Otherwise, coarse roughening would make it many times larger due to the increased surface area exposed to the field. Furthermore, we tried to remove the rough surface from Sample 2-R by cutting the edges as clean as possible. The cleaned Sample 2-R-C showed that $A$ has decreased by 0.1 nm/T$^{2.3}$ from that of Sample 2-R. This re-enforces our conclusion that surface roughness is not a dominant source determining $A$ and plays only minor role. We also conclude that the explanation proposed to reconcile TDR and $H_{c1}$ measurements \cite{Klein} does not hold the ground. Instead, we think that the field of first penetration measured by the Hall-probe was determined by the surface barrier rather than $H_{c1}$. Finally, we conclude that global measurements provide a reliable estimate of the overall behavior of the London penetration depth by sampling all sample surfaces in contrast to the local probes that are affected by the surface topography on the mesoscopic scale.

\begin{figure}[tb]
\includegraphics[width=8.5cm]{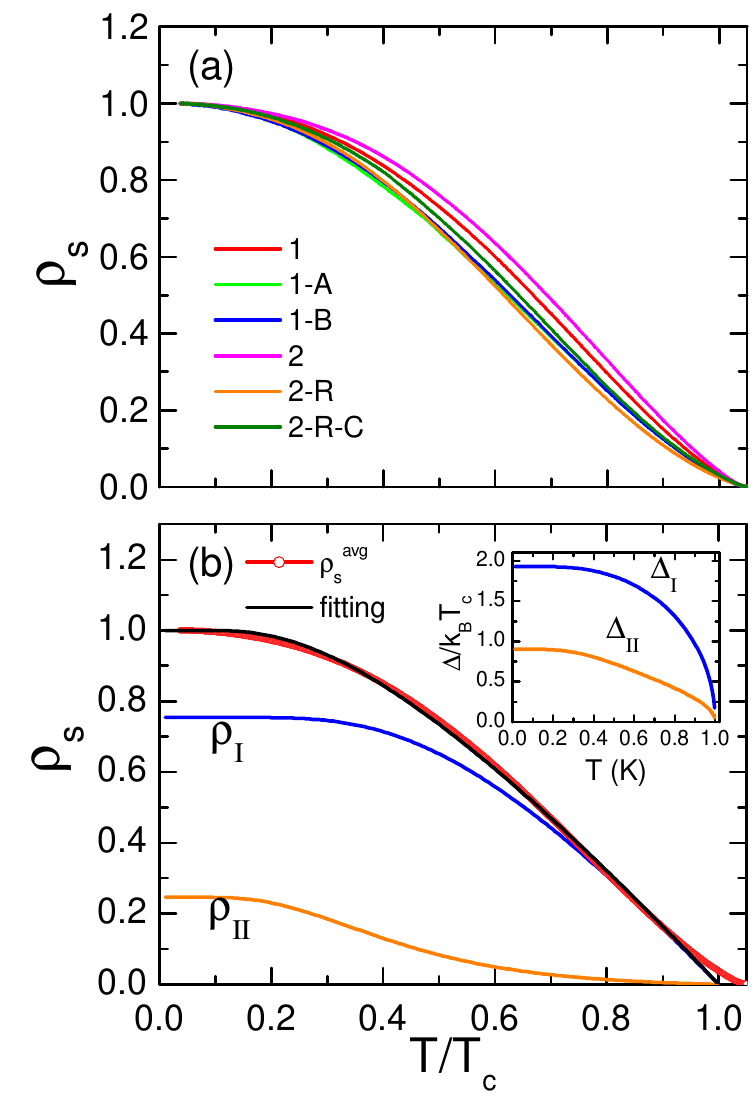}
\caption{\label{fig.5} (Color online) Superfluid density $\rho _{s} (T) = (\lambda(0)/\lambda(T))^2$ versus $T/T_c$. $\lambda(0)$ = 560 nm obtained from the previous report \cite{Kim2}. (a) $\rho_{s} (T)$ for all six samples. (b) The average superfluid ndensity for all six samples, $\rho_s ^{avg}$ (Red circle), is fitted to the two-gap $\gamma$ model \cite{Kogan}. Black solid line is the fitting result with $\rho = \gamma \rho _{_I} + (1- \gamma) \rho _{_{II}}$. The inset shows two gaps acquired from the fit, $\Delta _{_I}$ and $\Delta _{_{II}}$.}
\end{figure}

The superfluid densities $\rho _s = (\lambda (0) / \lambda (T))^2$ for all samples are shown in Fig.~\ref{fig.5}. The absolute penetration depth, $\lambda(0)=560$ nm was determined by TDR measurements of Al-coated sample \cite{Kim2}. The variation of $\rho_s$ among different samples is shown in Fig.~\ref{fig.5} (a). For the purpose of fitting, the average $\rho_{s}^{avg}$ is calculated from $\rho_{s}$ of samples 1 and 2 which have lowest $A$ values. The fitting was done with a self-consistent two-gap $\gamma$-model, where two gaps are calculated self-consistently at each temperature and at each iteration. The total superfluid density is given by $\rho = \gamma \rho_{_I} + (1- \gamma) \rho_{_{II}}$ \cite{Kogan}. The partial densities of states are chosen to be equal on the two bands, $n_1 = n_2 = 0.5$ and Debye temperature of 230 K was used to calculate the experimentally observed $T_c = 14.8$ K, which fixes the coupling constants (we used $\lambda_{11}$). Figure~\ref{fig.5}(b) demonstrates a good agreement between experimental $\rho_{s}^{avg}$ (symbols) and fitting (black solid line). The parameters acquired from the fit are: $\lambda_{11}$ = 0.66, $\lambda_{22}$ = 0.44, $\lambda_{12}$ = 0.07, $\lambda_{eff}$ = 0.34, $\gamma$ = 0.75 and $T_c$ = 14.95 K. This result indicates that 75 \% contribution of superfluid density comes from the band with $\rho_{_I}$ which has the larger gap $\Delta_{_I}$. The inset in Fig.~\ref{fig.5} (b) shows the behaviors of two energy gaps $\Delta_{_I}$ and $\Delta_{_{II}}$ versus temperature. Clearly, the smaller gap has significantly non-BCS temperature dependence. The zero temperature energy gaps $\Delta_{_I} (0)$ and $\Delta_{_{II}} (0)$ are 2.5 meV ($\Delta_{_I} (0) / k_{B} T_{c}$ = 1.93) and 1.1 meV ($\Delta_{_{II}} (0)/ k_{B} T_{c}$ = 0.9) respectively. The superconducting energy gaps  has also been probed in different experiments. From $\mu$SR \cite{Biswas2010, Bendele2010} and penetration depth \cite{Kim2} measurements, two s-wave energy gaps were reported with gap values similar to our results. $\mu$SR studies in FeSe$_{0.5}$Te$_{0.5}$ \cite{Biswas2010, Bendele2010} revealed two gaps of $\Delta_{large}$ $\sim$ 2.6 meV and $\Delta_{small} \sim$ 0.5-0.87 meV, and the penetration depth study \cite{Kim2} also showed that $\Delta_{large}$ $\sim$ 2.1 meV and $\Delta_{s} \sim$ 1.2 meV. According to scanning tunneling spectroscopy study, only one s-wave gap $\Delta \sim 2.3$ meV was observed in FeSe$_{0.4}$Te$_{0.6}$ \cite{Kato2009}, which is similar to the large gap $\Delta_{I} (0)$ of our result. However, rather large single or multi-gaps were reported from specific heat \cite{Hu2011}, optical conductivity \cite{Homes2010}, point-contact Andreeve reflectivity \cite{Park2010}, and angle-resolved photoemission spectroscopy \cite{Nakayama2010} suggesting strong-coupling superconductivity. The electronic specific heat in Fe(Te$_{0.57}$Se$_{0.43}$) \cite{Hu2011} revealed two energy gaps with $\Delta_{large} \sim$ 7.4 meV and $\Delta_{small} \sim$ 5.0 meV. From the optical conductivity in FeTe$_{0.55}$Se$_{0.45}$, two large energy gaps were also found with $\Delta_{large} \sim$ 5.1 meV and $\Delta_{small} \sim$ 2.5 meV. The point-contact Andreev reflectivity in FeTe$_{0.55}$Se$_{0.45}$ is consistent with single gap s-wave symmetry with $\Delta$ (at 1.70 K) $\sim$ 3.8 meV. Angle-resolved photoemission spectroscopy in FeTe$_{0.7}$Se$_{0.3}$ \cite{Nakayama2010}, an s-wave single gap of $\Delta \sim$ 4 meV was also observed. Overall, the pairing symmetry in FeTe$_{1-x}$Se$_{x}$ is still under debate.

To summarize, London penetration depth was measured in optimally-doped single-crystals of Fe(Te$_{0.58}$Se$_{0.42}$) of different size, shape and surface conditions to separate extrinsic and intrinsic effects determining the temperature variation of $\lambda(T)$. Even though there is a natural dispersion among six different samples, the average exponent $n_{avg}$ and pre-factor $A_{avg}$ are found to be 2.3 $\pm$ 0.1 and 1.0 $\pm$ 0.1 nm/K$^{2.3}$. The two-band $\gamma$ model fits the superfluid density rather well resulting in $\Delta_{_{I}}(0)$/$k_B T_c$ = 1.93 and $\Delta_{_{II}}(0)$/$k_B T_c$ = 0.9. From comparison of six samples, it is found that there exists unavoidable micro(meso)-scopic surface roughness on the sample edges and the effect of the roughness on pre-factor $A$ becomes more important as the sample size decreases, whereas the exponent $n$ remains practically unchanged. Overall, we conclude that global measurements provide an accurate and objective way to determine London penetration depth in pnictide superconductors.

\begin{acknowledgments}
The work at Ames was supported by the U.S. Department of Energy, Office of Basic Energy Sciences, Division of Materials Sciences and Engineering under contract No. DE-AC02-07CH11358. The work at Tulane was supported by the NSF under grants DMR-0645305 and EPS-1003897.
\end{acknowledgments}


\begin{thebibliography}{99}
\bibitem{Klein} T. Klein \emph{et al.}, \prb {\bf 82}, 184506 (2010).
\bibitem{Hsu} F. C. Hsu \emph{et al.}, Proc. Natl. Acad. Sci. {\bf 105}, 14262 (2008).
\bibitem{Kamihara} Y. Kamihara \emph{et al.}, J. Am. Chem. Soc. {\bf 128}, 10012 (2006).
\bibitem{Rotter} M. Rotter \emph{et al.}, \prl {\bf 101}, 107006 (2008).
\bibitem{Ogino} H. Ogino \emph{et al.}, Supercond. Sci. Technol. {\bf 22}, 075008 (2009).
\bibitem{Zhu} X. Zhu \emph{et al.}, \prb {\bf 79}, 024516 (2009).
\bibitem{Wang} X. C. Wang \emph{et al.}, Solid State Comm. {\bf 148}, 538 (2008).
\bibitem{Yeh} K. W. Yeh \emph{et al.}, Europhys. Letts. {\bf 84}, 37002 (2008).
\bibitem{Mizuguchi} Y. Mizuguchi \emph{et al.}, J. Phys. Soc. Japan {\bf 78}, 074712 (2009).
\bibitem{Prozorov} R. Prozorov and R. W. Giannetta, Supercond. Sci. Technol. {\bf 19}, R41 (2006).
\bibitem{Bendele} M. Bendele \emph{et al.}, \prb {\bf 82}, 212504 (2010).
\bibitem{Fang} M. H. Fang \emph{et al.}, \prb {\bf 78}, 224503 (2008).
\bibitem{Liu2009}T.~J. Liu \emph{et al.}, \prb {\bf 80}, 174509 (2009).
\bibitem{Khasanov} R. Khasanov \emph{et al.}, \prb {\bf 80}, 140511(R) (2009).
\bibitem{Margadonna} S. Margadonna \emph{et al.}, \prb {\bf 80}, 064506 (2009).
\bibitem{Sales} B. C. Sales \emph{et al.}, \prb {\bf 79}, 094521 (2009).
\bibitem{Subedi} A. Subedi \emph{et al.}, \prb {\bf 78}, 134514 (2008).
\bibitem{Xia} T. L. Xia \emph{et al.}, \prb {\bf 79}, 140510 (2009).
\bibitem{Bao} W. Bao \emph{et al.}, \prl {\bf 102}, 247001 (2009).
\bibitem{Qiu} Y. Qiu \emph{et al.}, \prl {\bf 103}, 067008 (2009).
\bibitem{Kotegawa} H. Kotegawa \emph{et al.}, J. Phys. Soc. Jpn. {\bf 77}, 113703 (2008).
\bibitem{Dong} J. K. Dong \emph{et al.}, \prb {\bf 80}, 024518 (2009).
\bibitem{Khasanov2} R. Khasanov \emph{et al.}, \prb {\bf 78}, 220510(R) (2008).
\bibitem{Kim2} H. Kim \emph{et al.}, \prb {\bf 81}, 180503(R) (2010).
\bibitem{Serafin} A. Serafin \emph{et al.}, \prb {\bf 82}, 104514 (2010).
\bibitem{Van Degrift} T. Van Degrift \emph{et al.}, Rev. Sci. Instrum. {\bf 46}, 599 (1975).
\bibitem{Prozorov2} R. Prozorov \emph{et al.}, \prb {\bf 62}, 115 (2000).
\bibitem{Hardy93} W. N. Hardy \emph{et al.}, \prl {\bf 70}, 3999 (1993).
\bibitem{Kogan} V. G. Kogan \emph{et al.}, \prb {\bf 80}, 014507 (2009).
\bibitem{Biswas2010} P. K. Biswas \emph{et al.}, \prb {\bf 81}, 092510 (2010).
\bibitem{Bendele2010} M. Bendele \emph{et al.}, \prb {\bf 81},
224520 (2010).
\bibitem{Kato2009} T. Kato \emph{et al.}, \prb {\bf 80}, 180507(R) (2009).
\bibitem{Hu2011} J. Hu \emph{et al.}, \prb {\bf 83}, 134521 (2011).

\bibitem{Homes2010} C. C. Homes \emph{et al.}, \prb {\bf 81}, 180508(R) (2010).

\bibitem{Park2010} W. K. Park \emph{et al.}, eprint  arXiv:1005.0190.

\bibitem{Nakayama2010} K. Nakayama \emph{et al.}, \prl {\bf 105}, 197001 (2010).
\end{thebibliography}
\end{document}